\documentclass[10pt]{iopart}
\makeatletter
\renewenvironment{abstract}
  {\section*{\abstractname}}
  {}
\makeatother

\usepackage[backend=biber,language=auto,autolang=other, style=phys]{biblatex}  
\AtEveryBibitem{%
  \clearfield{language}%
  \clearlist{language}%
}
\usepackage{amssymb}
\expandafter\let\csname equation*\endcsname\relax
\expandafter\let\csname endequation*\endcsname\relax
\usepackage{amsmath, amsbsy}
\usepackage{graphicx}
\usepackage[unicode=true, bookmarks=true, bookmarksnumbered=true, bookmarksopen=true, bookmarksopenlevel=1, breaklinks=false,pdfborder={0 0 0}, backref=false, colorlinks=true]{hyperref}
\hypersetup{citecolor=blue,linkcolor=blue}
\usepackage{breakurl}
\usepackage[dvipsnames]{xcolor} 
\usepackage{lipsum,multicol} 
\usepackage [english]{babel}
\usepackage [autostyle, english = american]{csquotes}
\MakeOuterQuote{"}
\usepackage{subcaption}
\usepackage{graphicx}

\usepackage[commandnameprefix=always]{changes}
\usepackage{comment}
\usepackage{indentfirst}

\begin{document}

\title[]{Demonstration of H-mode Error Field Identification in a Single Discharge via Island Healing}

\author{
E.M. Bursch$^{1,2}$,
A. Xie$^{1,2}$,
J.L. Barr$^{3}$,
S.K. Kim$^{4}$,
N.C. Logan$^{1,2}$,
S.M. Yang$^{4}$,
J.G. Bak$^{5}$,
W. Choi$^{3}$,
Q. Hu$^{4}$,
J.W. Juhn$^{5}$,
J. Kim$^{5}$,
J. Lee$^{5}$,
N. Leuthold$^{1,2}$,
J.K. Park$^{6}$,
C. Paz-Soldan$^{1,2}$,
L. Piron$^{7,8}$,
G. Seo$^{6}$,
J. Seo$^{9}$
}

\address{
$^1$ Columbia University, New York, NY, United States of America \\
$^2$ Columbia Fusion Research Center, New York, NY, United States of America \\
$^3$ General Atomics, San Diego, CA, United States of America \\
$^4$ Princeton Plasma Physics Laboratory, Princeton, NJ, United States of America \\
$^5$ Korea Institute of Fusion Energy, Daejeon, Republic of Korea \\
$^6$ Seoul National University, Seoul, Republic of Korea \\
$^7$ University of Padova, Padova, Italy \\
$^8$ 2 Consorzio RFX (CNR, ENEA, INFN, Università di Padova, Acciaierie Venete SpA), Padova, Italy \\
$^9$ Chung-Ang University, Seoul, Republic of Korea \\
}

\begin{center}
\vspace{0.6cm}
\textsuperscript{*} E-mail of the corresponding author:  emb2333@columbia.edu\\
\end{center}
\vspace{10pt}
\begin{abstract}
\noindent Non-disruptive compass scan error field identification via island healing is demonstrated on KSTAR in a single H-mode discharge. This method can allow for up to a 75\% reduction in the operational time necessary for a complete compass scan by reducing the required discharges from four to one. Previous experiments on DIII-D, JET, and MAST-U have demonstrated the technique in Ohmic and L-mode but required multiple discharges and high fidelity real time $n=1$ magnetic diagnostics located close to the plasma, which are unlikely to be viable for future devices. The results presented here were made possible by implementing a new set of triggering algorithms relevant to fusion pilot plant operation into the KSTAR plasma control system, along with KSTAR's strong RMPs and long pulse lengths. This represents a significant step towards the disruption-free error field identification method being demonstrated for deployment on future disruption-averse tokamaks, including ITER and fusion pilot plants. Steps needed to close remaining gaps to viability are addressed.

\end{abstract}

\maketitle
\ioptwocol

\section{\label{sec:Back} Background}
Intrinsic error fields (EFs) can cause tearing modes, locked modes, and disruptions in tokamaks \cite{LaHaye1992,buttery_error_2000,Menard2010ProgressPlasmas,logan_empirical_2020,burschImprovedN1Empirical2026}. Even EFs too small to penetrate the rational surface may still degrade confinement and rotation \cite{Callen2011}. Therefore, avoiding their effects is essential for high performance tokamak scenarios, in particular those operating with strong non-axisymmetric fields (ex: for resonant magnetic perturbation (RMP) edge-localized mode (ELM) suppression \cite{Evans2006,Nazikian2015,Lunia_PPCF_2026}).

To mitigate EFs, tokamaks are built with sets of error field correction coils (EFCCs) designed to cancel out the intrinsic EF, but the problem of how to optimally correct the EF is nontrivial. There are a number of considerations, but the main three are: (1) that the EF is not constant across operating regimes, (2) the plasma response must be considered for effective correction, and (3) there must be a suitable way to identify the necessary EFCC currents for a given EF that is compatible with a fusion pilot plant (FPP).

A given tokamak has an EF composed of a set of static and dynamic sources. This can have strong effects on EF correction, even to the point that using EFCC currents based on a given L-mode (Low-confinement mode) scenario in an H-mode (High-confinement mode) scenario can increase the resultant error field to a level higher than if no correction was attempted at all \cite{pharr_error_2024, Piron_JET_EF_Control}. The important role of dynamic error fields is also highlighted by the increased level of EF found in KSTAR across operating points, as compared with the extremely low level identified in 2015 using an Ohmic compass scan \cite{In2015,Seo_Revised_KSTAR_EF}.

Of highest practical importance is the question of how to select the currents employed in a given set of EFCCs. There have been numerous techniques used to identify intrinsic EF, for which EFCC currents are then designed to compensate. The compass scan \cite{Hender1992EffectDischarges,LaHaye1992} is accomplished by ramping EFCCs in different toroidal phases and fitting their amplitude and phase to a circle, where the center is the amplitude and phase of the EF. The island steering method \cite{Strait2014,Shiraki2014,Shiraki2015,Strait_PPCF2025,Jiang_NF2026} uses the EFCCs to move a magnetic island across varying coil currents and finds the EF using a torque balance model. While this method is attractive for its time-dependent demonstration across H-mode and L-mode plasmas without disruptions, at reactor-relevant conditions with low $q_{95}$, moving a magnetic island through the plasma is likely untenable. Other methods vary coil current and identify the EF based on the rotation braking \cite{paz-soldan_spectral_2014,Reimerdes2009} or angular momentum optimization \cite{Lanctot2016}. A more detailed description of other methods can be found in Reference \cite{paz-soldan_non-disruptive_2022} and the references therein.

Of the leading methods, the most widely used is the compass scan. It is the most straightforward measurement of the threshold for penetration, which is the primary desired quantity in error field identification experiments. Also, it has been demonstrated that even compass scans with relatively large centroid uncertainties can yield EF correction settings that substantially improve plasma performance \cite{pharr_error_2024,}. However, traditionally, the compass scan has been accomplished disruptively, with each coil current ramp for the given toroidal phase triggering error field penetration, a locked mode, and a disruption. Given that one of the principal goals of correcting EFs is avoiding disruptions, any method of EF identification which is necessarily disruptive is unsuitable. Fortunately, as first demonstrated on DIII-D \cite{paz-soldan_non-disruptive_2022,hu_non-disruptive_2024}, as well as JET and MAST-U \cite{Piron_JET, piron_locked_2023, Piron_NF2024}, it is possible to conduct a compass scan non-disruptively via magnetic island healing. 

By implementing a real-time control algorithm based on detecting the locked mode onset, control action can be taken to remove the RMPs and puff gas to heal the magnetic islands prior to disruption. The technique has proved successful and portable, making it an attractive candidate for disruption-averse tokamaks such as ITER and FPPs. 
Despite its successful demonstrations, there are several remaining limitations: it requires significant operational time (it has only been accomplished with one or two RMP ramps), it requires diagnostics likely not relevant to FPPs, and it has only been done in Ohmic and L-mode plasmas (despite H-mode being the most critical operating regime for FPP relevance).

This work shows that each of these limitations can be adequately addressed. The experiments presented here accomplish a full compass scan in one discharge without disrupting, used FPP-relevant diagnostics, and conducted non-disruptive scans in H-mode.

\section{\label{sec:coil_conv} Representation of RMP Currents}
\begin{figure*}[t]
  \centering
  \includegraphics[width=.99\linewidth]{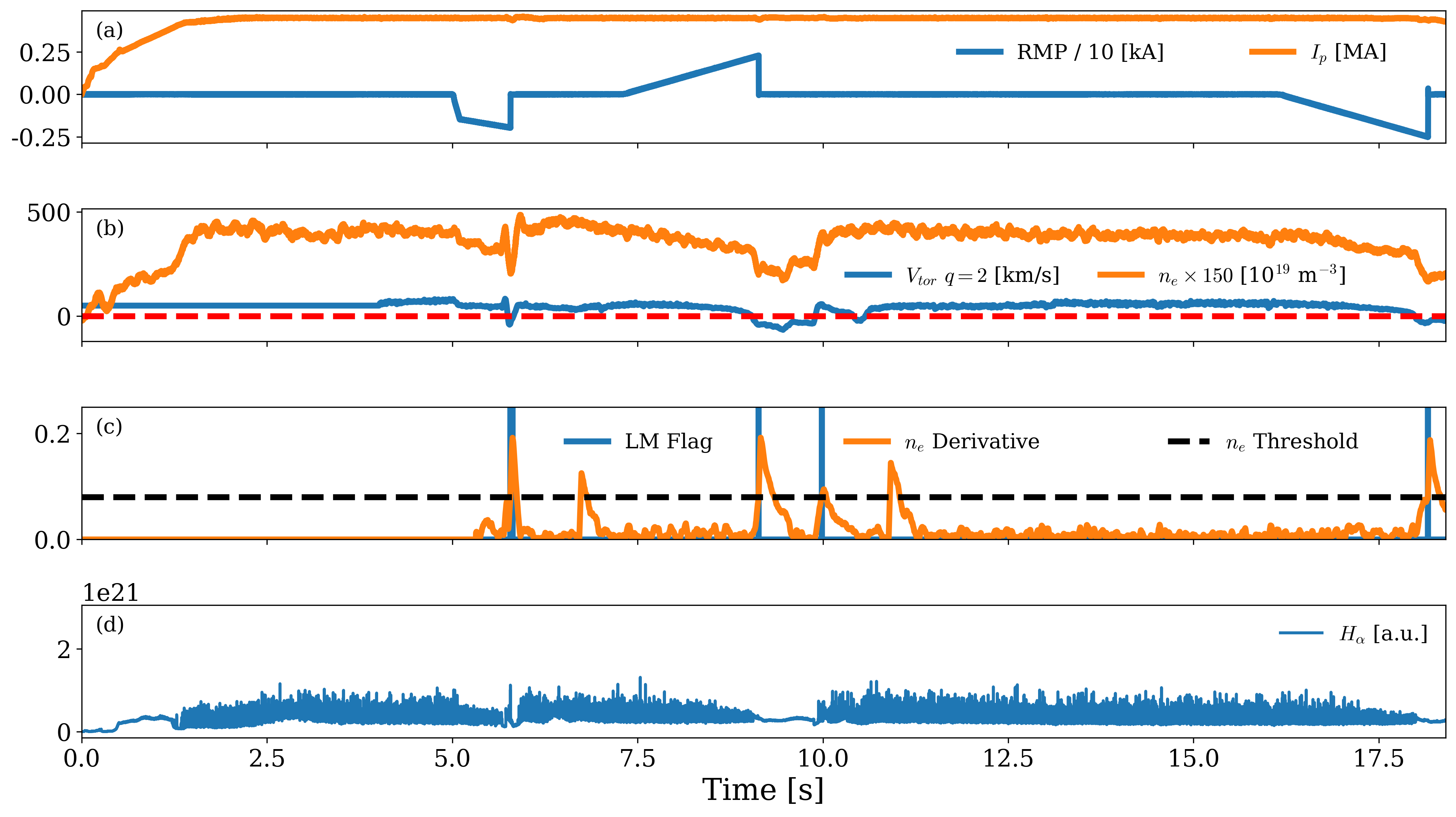}
  \caption{Time trace showing the single shot non-disruptive compass scan completed in KSTAR shot 42485. Panel a) is plasma current and RMP coil current. Panel b) is the $q=2$ toroidal rotation and the line-averaged density. Panel c) is the derivative of the window averaged real-time $n_e$ signal, along with the triggering threshold, and the locked mode flag sent to the PCS. Panel d) is the raw $H_\alpha$ signal.}
  \label{fig:3ramps_timetrace}
\end{figure*}
The results of the EF identification (both traditional and non-disruptive) will be presented in terms of $n=1$ RMP strength and the core dominant mode overlap ($\delta$). For results using the $n=1$ RMP strength, the magnitude and direction of the $n=1$ perturbation from the midplane are used to represent the given configuration. These experiments used equal amplitude currents in all three rows of KSTAR's RMP coils with a 90-degree phasing between each row. Unless otherwise noted, the representative amplitude and phase of the $n=1$ waveform applied in the midplane coil is used here to denote the scan configuration.

The other representation is the core dominant mode overlap, $\delta$, which quantifies resemblance of the 3D perturbation to the most dangerous spectrum. It is calculated here using the Generalized Perturbed Equilibrium Code (GPEC) \cite{park_computation_2007}. More details on the derivation and explanation of $\delta$ can be found in Reference \cite{pharrCoordinateInvariant2026}. This method is directly comparable across RMP configurations regardless of how many rows of coils or what phase between each row is selected. For that reason, the figures presented here will use $\delta$, but the intrinsic EFs identified from each compass scan will be reported in text using both the $n=1$ and $\delta$ conventions.

In addition, both the $n=1$ and the dominant mode overlap representations are normalized by a dimensionless line-averaged density at the time of error field penetration. This is done as:
\begin{equation}
    \bar{\delta}(t_\text{pen}) = \delta \times(\bar{n}_e/n_e(t_\text{pen}))
\end{equation}
Where $\bar{\delta}(t_\text{pen})$ is the core dominant mode overlap at the time of mode penetration, $\bar{n}_e$ is the mean of all the line-averaged densities at the time of penetration in the compass scan, and $n_e(t_\text{pen})$ is the line-averaged density at the time of mode penetration. Since density has been found to play a large role in penetration thresholds \cite{Buttery1999,Wang2018DensityEAST,Wang2014StudyJ-TEXT}, this normalization is often done to prevent density fluctuations obscuring the error field penetration results. As in Ref. \cite{Seo_Revised_KSTAR_EF}, the local error field penetration ($t_\text{pen}$) used for the compass scans is identified from $q=2$ electron cyclotron emission. 

\section{Single Discharge Compass Scan Demonstration}
As detailed in Section \ref{sec:alg}, new FPP-relevant locked mode detection algorithms were implemented on KSTAR. In KSTAR shot 42485, these were used to complete three RMP ramps during the $I_p$ flattop without disrupting the plasma. This provides enough error field penetration points to fit a circle and identify an EF. A full four point scan (allowing for uncertainty quantification) is possible with the optimized settings found during the experiment. Four ramps were attempted, but only three were executed due to the PCS settings producing a false positive flag in the density algorithm. In Figure \ref{fig:3ramps_timetrace}, each of the RMP ramps can be seen, followed by mode penetration, a decrease to zero of the $q=2$ plasma rotation, an H-L back-transition in the $H_\alpha$, and then recovery. Although minor perturbations to the $I_p$ are observed, disruptions are successfully avoided throughout the full shot duration. For reference, one of the ramps from Figure \ref{fig:3ramps_timetrace} is compared with a disruptive ramp in Figure \ref{fig:disrupt_comp} of the Appendix.

\begin{figure}[h]
  \centering
  \includegraphics[width=.8\linewidth]{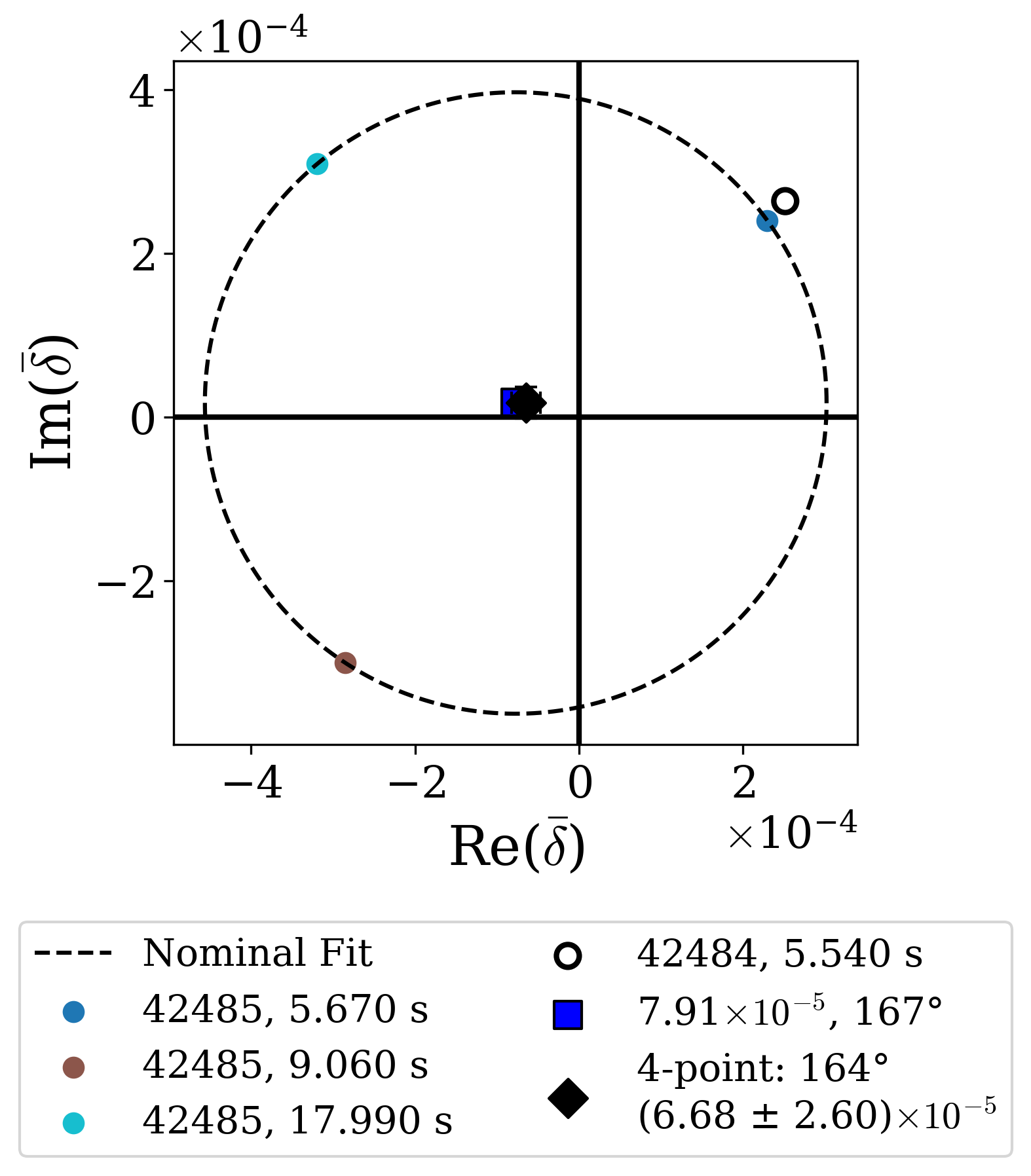}
  \caption{Single discharge density-normalized $\bar{\delta}$ compass scan at 1.9 T $B_T$, 0.45 MA $I_p$, and $q_{95} \approx 6$. The open circle is the only other EF ramp done at identical plasma conditions.}
  \label{fig:HModeSingleScan}
\end{figure}

Three point circle fitting yields a deterministic center but cannot generate covariance based uncertainty bounds; consistency with an independent separate shot penetration point increases confidence in the derived EF centroid. The circle, with a center at a $\bar{\delta}$ of $7.91\times 10^{-5}$ and $167^\circ$ is plotted in Figure \ref{fig:HModeSingleScan}. This is equivalent to an $n=1$ RMP with $586$ Amps (A) and a phase of $103^\circ$ (assuming equal amplitude and $90^\circ$ phasing). For reference, the only other data point collected under the same plasma parameters is also plotted on the figure.
While a full comparison of the identified EF using the single discharge disruption-free method cannot be undertaken due to different operating points, the additional data point collected (represented by the open circle in Figure \ref{fig:HModeSingleScan}), lies extremely close to the circle fit from the single shot. Finding the centroid of the circle from all four points results in $\bar{\delta}$ of $(6.68 \pm 2.60)\times 10^{-5}$ and $164^\circ$. This is equivalent to an $n=1$ RMP with $494 \pm 130$  A and a phase of $106^\circ$. Therefore, the single discharge compass scan centroid successfully falls within the error bars of the combined compass scan.

\section{\label{sec:alg} Detection Algorithms}
The algorithm implemented in the KSTAR plasma control system (PCS) to enable these experiments consists of 2 branches. The Ohmic/L-mode branch relies on a real-time density algorithm and a real-time $H_\alpha$ algorithm ($H_\alpha$ measures light emitted by excited Hydrogen atoms). The $H_\alpha$ signal increase is a result of particles being expelled from the plasma following the local penetration, during the global locking event. The H-mode branch relies on the same real-time density algorithm as well as a real-time dZ/dt algorithm (where dZ/dt denotes the derivative in the plasma vertical position). In each mode, if either algorithm raises a flag, a control action is taken. Previous non-disruptive error field experiments relied on strong $n=1$ magnetic field signals \cite{paz-soldan_non-disruptive_2022,Piron_JET,hu_non-disruptive_2024} to trigger their event response, but FPP operation likely precludes such techniques \cite{Piron_NF2024}. The specific input signals used on KSTAR were chosen both for their relevance to ITER and FPP operation and because similar magnetic triggering was not possible. Density and dZ/dt are planned for ITER and FPP operation \cite{campbell_introduction_2025,reinke_overview_2024,biel_diagnostics_2019}, and $H_\alpha$ is planned for ITER \cite{asadulin_tunable_2021} but is more uncertain for FPPs.

For each algorithm, the raw signal is first smoothed using a window averaging technique. Then, the derivative of the averaged window is calculated in real-time, starting when the RMP coils begin to ramp (to avoid any false triggering prior to the resonant magnetic perturbation). When the derivative of the averaged signal surpasses a user-specified threshold, the PCS raises a flag that is passed to the Off Normal Fault Response (ONFR) system. Then, the ONFR is able to turn off the RMP, puff gas as desired, and then start a new RMP ramp with a new phase after a predetermined wait time, allowing the plasma to heal before the next RMP ramp. 

\begin{figure}[h]
  \centering
  \includegraphics[width=1\linewidth]{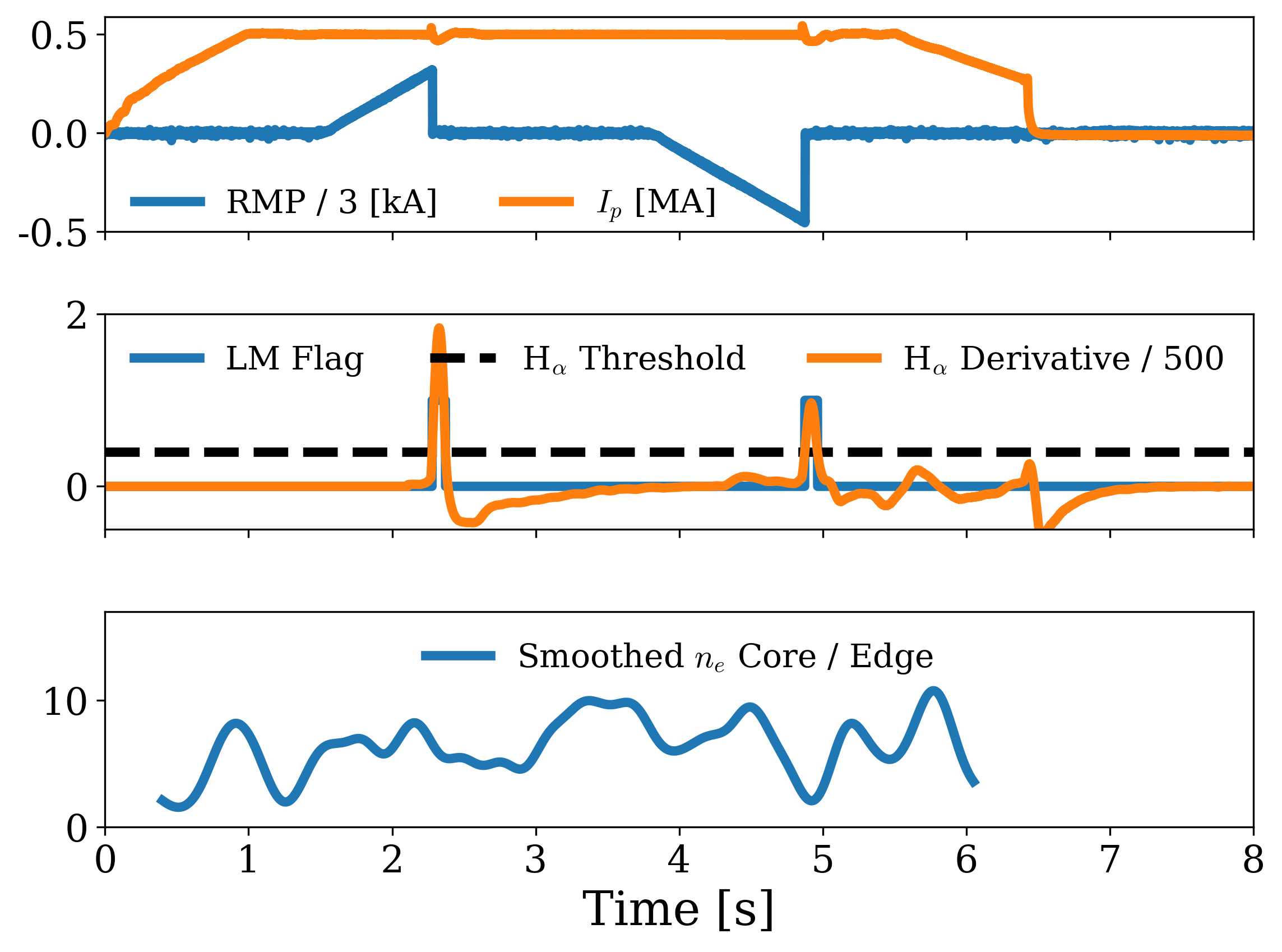}
  \caption{Time traces of KSTAR shot 42476. The top panel is plasma current and RMP coil current. The middle panel is the derivative of the window averaged real-time $H_\alpha$ signal, along with the triggering threshold, and the locked mode flag sent to the PCS. The bottom panel is the smoothed ratio of a core to an edge $n_e$ channel.}
  \label{fig:42476_ha_triggering}
\end{figure}

Figure \ref{fig:42476_ha_triggering} shows an example of triggering by the Ohmic $H_\alpha$ signal. As the RMP coil current is ramped up, the local error field penetration occurs, followed by a spike in $H_\alpha$ signal just before the global locked mode. In H-mode shots, where rotation data is available, this coincides with the toroidal rotation reaching zero. The control action is taken since the $H_\alpha$ signal threshold is surpassed, and the RMP current is set to zero. If desired, the control action can also include puffing in gas to help heal the island. This was found to be unnecessary in many shots due to the high $q_{95}$ operational point chosen. Similar progressions are seen for both the density algorithm and the dZ/dt algorithm implementations.

These algorithms, along with the magnetics-only approach on other devices, range in difficulty from an operational standpoint. They similarly range in delay from error field penetration to detection time. The quickest method is the most difficult operationally for ITER and FPPs as it requires clean, real-time $n=1$ magnetics signals close to the plasma. The methods presented here have larger delays but are simpler to implement from an operational standpoint.

\section{Ohmic Demonstration of the Control 
Algorithm}
\begin{figure}[h]
  \centering
  \includegraphics[width=.8\linewidth]{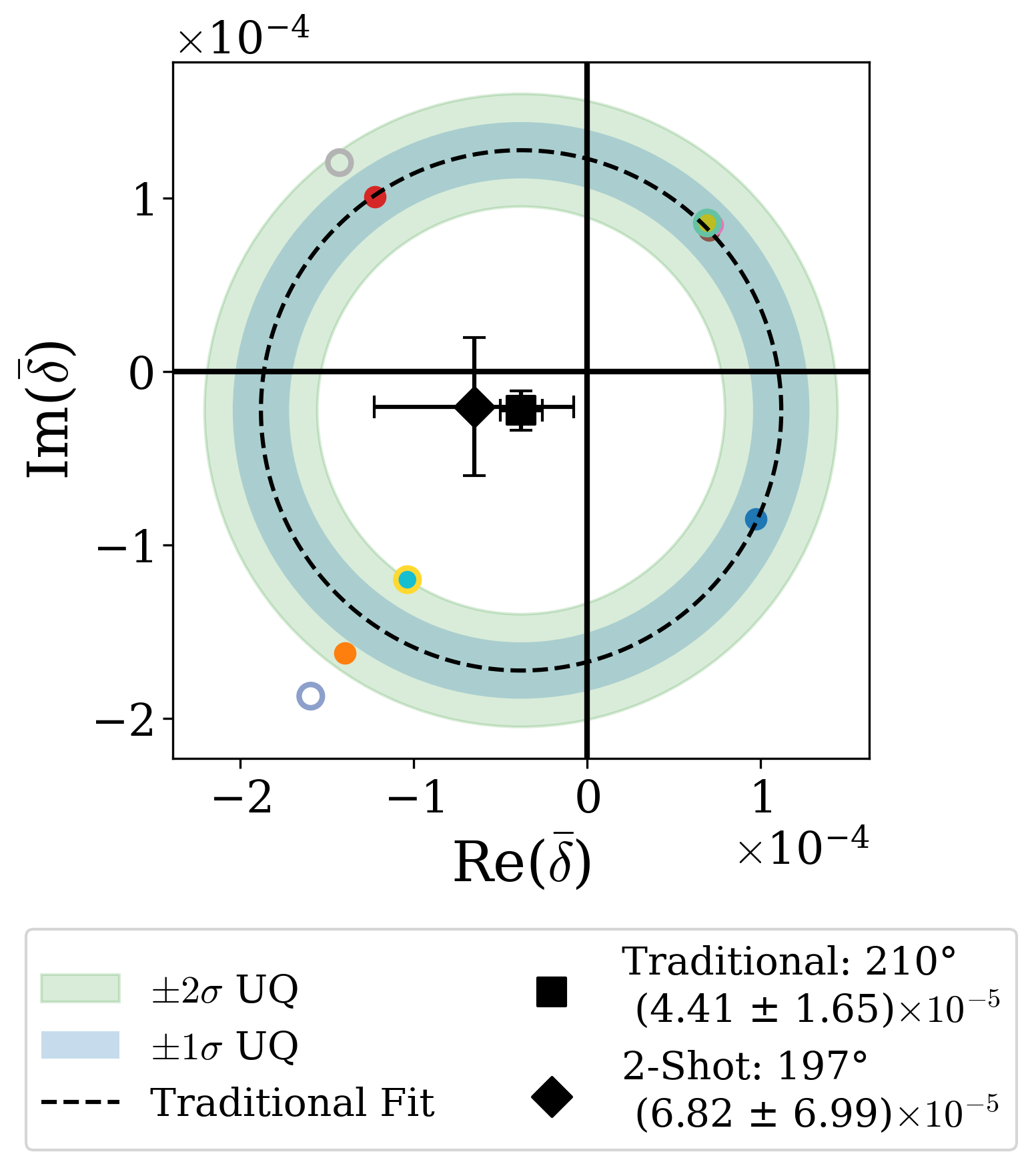}
  \caption{Combined density-normalized $\bar{\delta}$ compass scan including both traditional and non-disruptive Ohmic results. Solid points are from traditional compass scan RMP ramps and hollow points are from the 2-shot non-disruptive compass scan. The conditions included 1.8 T $B_T$, 0.5 MA $I_p$, and $q_{95} \approx 4.6$. Blue and green bands correspond to 1 and 2 $\sigma$ of uncertainty in the fitted circle from only the traditional compass scan points. Corresponding shots and times for the points can be found in the Appendix.}
  \label{fig:Ohmic_scan}
\end{figure}

The experimental results also successfully demonstrated the algorithm in Ohmic discharges. This brings the total number of devices with successful non-disruptive EF identification up to four (DIII-D, JET, MAST-U, and KSTAR), which lays a strong groundwork for extrapolation to ITER and FPPs. The Ohmic plasmas were operated at a toroidal field ($B_T$) of 1.8 Tesla (T), plasma current ($I_p$) of 0.5 MA, and a safety factor at 95\% of the normalized poloidal flux ($q_{95}$) near 4.6. Both the Ohmic and H-Mode $q_{95}$ values in these experiments were higher than those planned for ITER and most FPPs. However, non-disruptive compass scans have been done at lower $q_{95}$ on DIII-D \cite{hu_non-disruptive_2024} leading to a required actuation time of between 100 and 500 ms for ITER \cite{hu_non-disruptive_2024}. Enough data was collected during the experiment to produce a traditional compass scan (consisting of only the first RMP ramps) as well as an Ohmic compass scan completed in two shots (with two 3D ramps per shot). The results can be found in Figure \ref{fig:Ohmic_scan}.
The center of the fitted circle corresponds to the identified magnitude and direction of the intrinsic error field. The center of the traditional compass scan is at a $\bar{\delta}$ of $(4.41 \pm1.65) \times 10^{-5}$ and $210^\circ$. This is equivalent to an $n=1$ RMP with $140 \pm 38$  A and a phase of $63^\circ$ (assuming equal amplitude and $90^\circ$ phasing). The center of the non-disruptive, 2-shot compass scan is at a $\bar{\delta}$ of $(6.82 \pm6.99 )\times 10^{-5}$ and $197^\circ$. This is equivalent to an $n=1$ RMP with $267 \pm 214$  A and a phase of $78^\circ$. Although the identified EF is of similar magnitude to the most recent traditional EF identification \cite{Seo_Revised_KSTAR_EF}, is not directly comparable due to the difference in operating regime. This is because intrinsic EF magnitude and direction are often different for different plasmas due to differences in the currents in the various EF sources, particularly when comparing Ohmic to H-mode EFs \cite{pharr_error_2024}. 

As seen in Figure \ref{fig:Ohmic_scan}, the radii obtained from the second 3D RMP ramp in each discharge (the hollow points on the left half of the figure) are larger than the traditional compass scan points. This observation is consistent with previous observations on DIII-D \cite{paz-soldan_non-disruptive_2022}. The density profile peaking changes throughout the shot and so the density closest to the $q=2$ surface shifts relative to the line-averaged density. This trend can be seen in the bottom subplot of Figure \ref{fig:42476_ha_triggering}. However, the uncertainty in the measurements as well as the possibility of electromagnetic forces from the central solenoid biasing the measurements \cite{paz-soldan_non-disruptive_2022} makes a clear mechanism across the shots difficult to identify. At present we cannot definitively separate profile driven effects from solenoid induced bias, which motivates further dedicated experiments. Despite these differences, the centers of the traditional and non-disruptive scans are within uncertainty calculated from the covariance of the least-squares fit. The uncertainty could be reduced by including more points in the scan, particularly with different phasing directions. Therefore, the successful operation of the control algorithm across these shots demonstrates the suitability of the FPP-relevant locked mode detection in KSTAR Ohmic plasmas. 

\section{H-mode Demonstration of the Control Algorithm}
\begin{figure}[h]
  \centering
  \includegraphics[width=.8\linewidth]{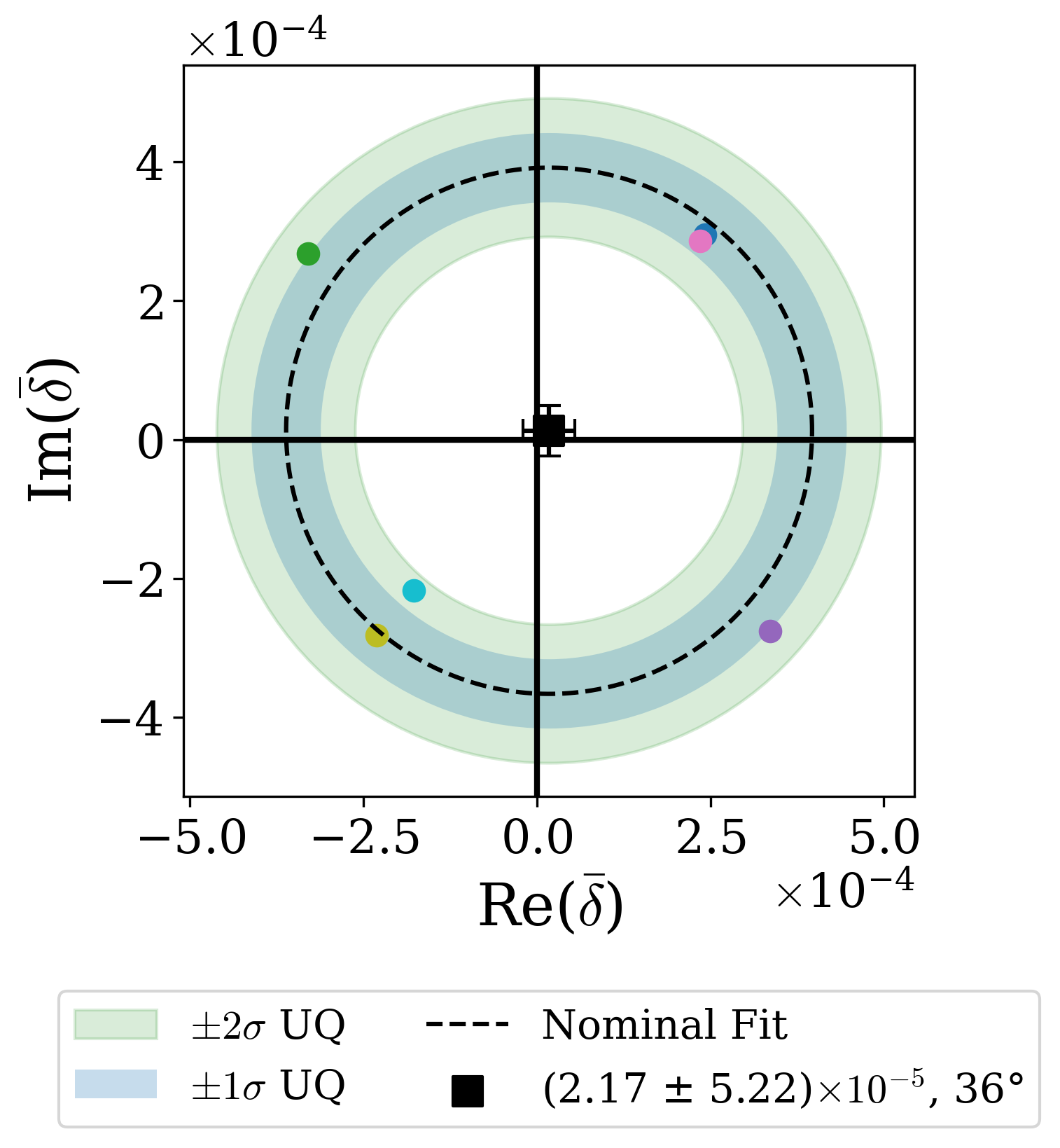}
  \caption{Traditional H-mode density-normalized $\bar{\delta}$ compass scan at 1.9 T $B_T$, 0.5 MA $I_p$, and $q_{95} \approx 5.6$. Blue and green bands correspond to 1 and 2 $\sigma$ of uncertainty in the fitted circle. Corresponding shots and times for the points can be found in the Appendix.}
  \label{fig:HModeTradScan}
\end{figure}

Prior disruption-free compass scan EF identification experiments have not been conducted in H-mode. However, H-mode operation poses the largest risk in terms of disruptions. In addition, EF correction found in Ohmic discharges and applied to H-mode can make the intrinsic EF worse \cite{pharr_error_2024}. Using the outlined detection methods, we demonstrated non-disruptive RMP ramps across several H-mode discharges. The first ramps from each of these are compiled into a traditional compass scan in Figure \ref{fig:HModeTradScan}. The center of the fitted circle is at a $\bar{\delta}$ of $(2.17 \pm5.22) \times 10^{-5}$ and $36^\circ$. This is equivalent to an $n=1$ RMP with $165 \pm 274$  A and a phase of $239^\circ$ (assuming equal amplitude and $90^\circ$ phasing). While the uncertainty in the EF measurement is large, it could be reduced by increasing the number of points as well as measuring in different phasing directions. Differences between shots in plasma parameters such as rotation and $\beta_n$ increase the scatter. Future work could find an optimal model for normalizing the measured EF by other plasma parameters, as already done here with density. Here as before, although these results are not directly comparable with the most recent KSTAR EF identification \cite{Seo_Revised_KSTAR_EF} due to differing operating points, we see a similar magnitude and direction to their experimental results, which were found in a 0.6 MA $I_p$ and 2.4 T $B_T$ H-mode.

\section{\label{sec:Conc} Conclusion}
This paper presents a demonstration of disruption-free tokamak error field identification in a single discharge. It was accomplished using a newly implemented, FPP-relevant detection algorithm. This same algorithm was also used to demonstrate H-mode non-disruptive compass scans and more broadly increase confidence in the technique's transferability across existing tokamaks and ultimately to ITER and FPPs.

These results do include several limitations that must be addressed prior to confident deployment on disruption-averse devices. While detection methods and pulse lengths will only continue to improve, the high $q_{95}$ used throughout these experiments must be brought down to more FPP-relevant conditions. Work has been done in this area outside of H-mode on DIII-D \cite{hu_non-disruptive_2024}, but it must be extended to higher performance plasmas and completed within a single shot. Another limitation is the device-specific nature of the detection method and particularly the triggering thresholds. Since ITER and FPPs will not have the benefit of large databases of past discharges from which to set thresholds, a device-agnostic triggering method must be implemented. One solution could be the production of synthetic data for device-independent detection. Finally, although dZ/dt and density diagnostics are likely to exist on ITER and FPPs \cite{campbell_introduction_2025,reinke_overview_2024,biel_diagnostics_2019}, $H_\alpha$ is planned for ITER \cite{asadulin_tunable_2021} but is more uncertain for FPPs.

While the core EF is of greatest importance for locked modes and disruptions, the efficient and successful application of edge RMPs for ELM suppression would benefit from this technique being used for edge-specific EF identification. In addition, the non-disruptive RMP ramps could provide a similar benefit for the identification of optimal 3D RMP configurations for ELM suppression. Traditionally, $n=1$ coil currents for suppression are identified by ramping the RMPs until suppression and eventually a locked-mode induced disruption. EF knowledge is critical, since the magnitude of the measured EF can be a significant fraction of the window for ELM suppression \cite{Lunia_PPCF_2026,Gu2019EdgeDIII-D}. Since ELM suppression and disruption avoidance are both highly desirable for ITER and FPPs, a robustly non-disruptive method for ELM suppression coil optimization could prove useful. 

The successful demonstration of disruption-free EF identification in a single discharge is an important step in advancing towards robust EF identification on future disruption-averse tokamaks. Continued research on this topic is essential for confident deployment on ITER and FPPs. This work demonstrates that H-mode disruption-free tokamak EF identification is possible within a single discharge using magnetic island healing, which is a critical step forward for disruption-free operation of ITER and FPPs.

\section*{\label{sec:Ack} Acknowledgments}
The authors would like to acknowledge the help of Heather Shen and Hyunsun Han in the course of this research.

Work supported by US DOE under DE-SC0021968, DE-SC0026404, DE-FC02-04ER54698, and DE-AC02-09CH11466. Work supported by the Government of the Republic of Korea through the Korea Institute of Fusion Energy (KFE), including the R\&D Program of “KSTAR Experimental Collaboration and Fusion Plasma Research (EN2601-17),” funded by the Korea Ministry of Science and ICT (MSIT), the “High Performance Tokamak Plasma Research \& Development” program (No. EN2601), and the National R\&D Program through the National Research Foundation of Korea (NRF) and KFE under Grant Nos. RS-2026-25547568 and CN-2602.

\printbibliography

\newpage
\begin{figure*}[t]
\section*{Appendix: Supplemental Data}
    \centering
    \includegraphics[width=0.6\linewidth]{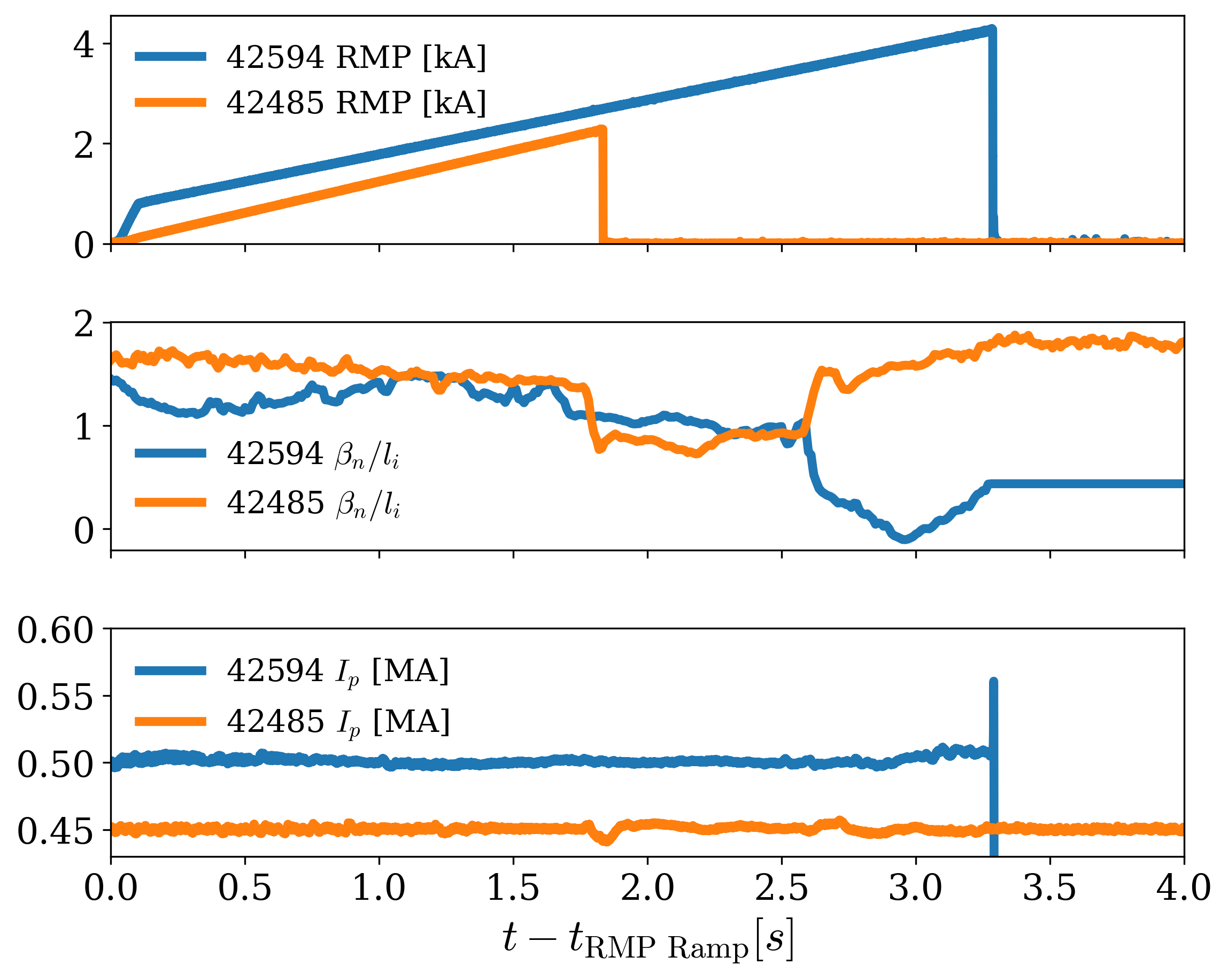}
    \caption{Comparison of KSTAR 42594 which disrupts during the RMP ramp and one of the RMP ramps from 42485, the single discharge compass scan. Although the plasmas were similar, 42594 contained a mix of $n=1$ and $n=2$ RMPs so without control intervention, shot 42485 would be expected to disrupt sooner than 42594.}
    \label{fig:disrupt_comp}
\end{figure*}
\begin{figure*}[htbp]
    \centering
    \makebox[\textwidth][c]{%
        \begin{subfigure}[t]{0.6\textwidth}
            \centering
            \includegraphics[width=\linewidth]{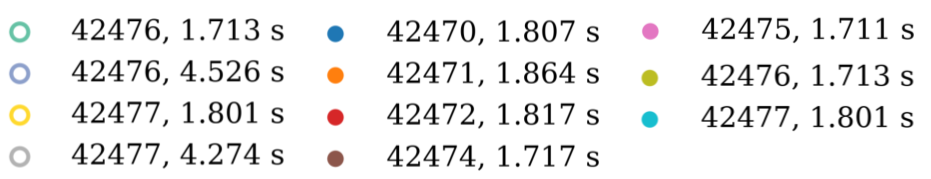}
            \caption{Shot and times for the points in Figure \ref{fig:Ohmic_scan}}
        \end{subfigure}
        \hspace{0.02\textwidth}
        \begin{subfigure}[t]{0.39\textwidth}
            \centering
            \includegraphics[width=\linewidth]{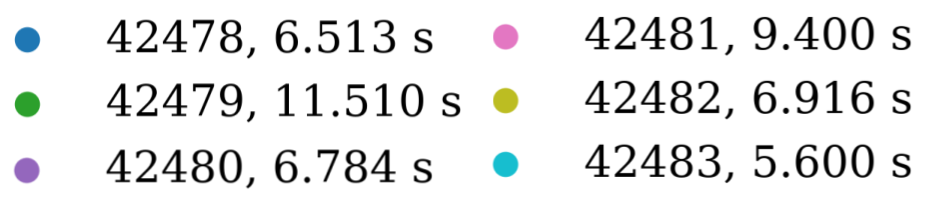}
            \caption{Shot and times for the points in Figure \ref{fig:HModeTradScan}}
        \end{subfigure}
    }

    \caption{%
        \parbox{\textwidth}{Shot and time information for the compass scans shown in Figures \ref{fig:Ohmic_scan} and \ref{fig:HModeTradScan}.}
    }
    \label{fig:comparison}
\end{figure*}

\begin{table*}
\centering
\label{tab:EF_centers}
\begin{tabular}{llccccc}
\hline
Compass Scan
& $B_T$ (T) & $I_p$ (MA)
& $n=1$ Amp.   A & $n=1$ Phase ($^\circ$)
& $\bar{\delta}$ Amp. & $\bar{\delta}$ Phase ($^\circ$) \\
\hline
O - T
& 1.8 & 0.5
& $140 \pm 38$  & $63$
& $(4.41 \pm 1.65)\times10^{-5}$ & $210$ \\

O - ND
& 1.8 & 0.5
& $267 \pm 214$ & $78$
& $(6.82 \pm 6.99)\times10^{-5}$ & $197$ \\

H - T
& 1.9 & 0.5
& $165 \pm 274$ & $239$
& $(2.17 \pm 5.22)\times10^{-5}$ & $36$ \\

H - ND
& 1.9 & 0.45
& $586$ & $103$
& $7.91\times10^{-5}$ & $167$ \\

H - ND+1
& 1.9 & 0.45
& $494 \pm 130$ & $106$
& $(6.68 \pm 2.60)\times10^{-5}$ & $164$ \\

\hline
\end{tabular}
\caption{Intrinsic EF centroids from each of the presented compass scan fits, along with the operating-point toroidal field and plasma current. O denotes Ohmic and H denotes H-mode. T denotes traditional compass scans and ND denotes non-disruptive compass scans. ND+1 is the H-Mode single shot compass scan with the one additional point at the same operating point to allow for uncertainty quantification.}
\end{table*}

\end{document}